\documentclass[twocolumn]{aastex62}

\usepackage{amsmath}
\usepackage{tikz}
\usepackage{graphicx}
\usepackage[normalem]{ulem}
\usetikzlibrary{positioning}
\usetikzlibrary{patterns}
\usetikzlibrary{decorations.pathmorphing}
\usetikzlibrary{arrows}

\newcommand{\rhogj}{\rho_\mathrm{co}}
\newcommand{\emin}{\epsilon_\mathrm{min}}
\newcommand{\temin}{\tilde{\epsilon}_\mathrm{min}}
\newcommand{\ellic}{\ell_\mathrm{IC}}
\newcommand{\ellgg}{\ell_{\gamma\gamma}}

\submitjournal{ApJL}

\shorttitle{Physics of Gaps in BH Magnetospheres}
\shortauthors{Chen et al.}

\begin{document}

\title{Physics of Pair Producing Gaps in Black Hole Magnetospheres}

\correspondingauthor{Alexander Y. Chen}
\email{alexc@astro.princeton.edu}

\author{Alexander Y. Chen}
\affil{Department of Astrophysical Sciences, Princeton University, Princeton, NJ 08544, USA}

\author{Yajie Yuan}
\altaffiliation{Lyman Spitzer, Jr. Postdoctoral Fellow}
\affiliation{Department of Astrophysical Sciences, Princeton University, Princeton, NJ 08544, USA}

\author{Huan Yang}
\affiliation{University of Guelph, Guelph, Ontario N2L 3G1, Canada}
\affiliation{Perimeter Institute for Theoretical Physics, Waterloo, Ontario N2L 2Y5, Canada}

\begin{abstract}
In some low-luminosity accreting supermassive black hole systems, the supply
of plasma in the funnel region can be a problem. It is believed that a local
region with unscreened electric field can exist in the black hole
magnetosphere, accelerating particles and producing high energy gamma-rays
that can create $e^{\pm}$ pairs. We carry out time-dependent self-consistent
1D PIC simulations of this process, including inverse Compton scattering and
photon tracking. We find a highly time-dependent solution where a macroscopic
gap opens quasi-periodically to create $e^{\pm}$ pairs and high energy
radiation. If this gap is operating at the base of the jet in M87, we expect
an intermittency on the order of a few $r_g/c$, which coincides with the time
scale of the observed TeV flares from the same object. For Sagittarius A* the
gap electric field can potentially grow to change the global magnetospheric
structure, which may explain the lack of a radio jet at the center of our
galaxy.
\end{abstract}


\section{Introduction}
\label{sec:intro}
Black holes can be powerful engines for active galactic nuclei (AGN), galactic
superluminal sources, gamma-ray bursts, and other energetic phenomena. It has been shown that the rotational energy of
a Kerr black hole can be electromagnetically extracted to launch powerful jets \citep{1977MNRAS.179..433B,2012MNRAS.423.3083M}. However, the
process relies on the field being frozen into the plasma, and if matter from the
accretion disk cannot easily cross the field line to enter the jet, some
mechanism of plasma supply in the jet funnel is needed
\citep[e.g.,][]{1977MNRAS.179..433B,1992SvA....36..642B,1998ApJ...497..563H}.

\begin{figure}
\centering
\includegraphics[width=0.95\columnwidth]{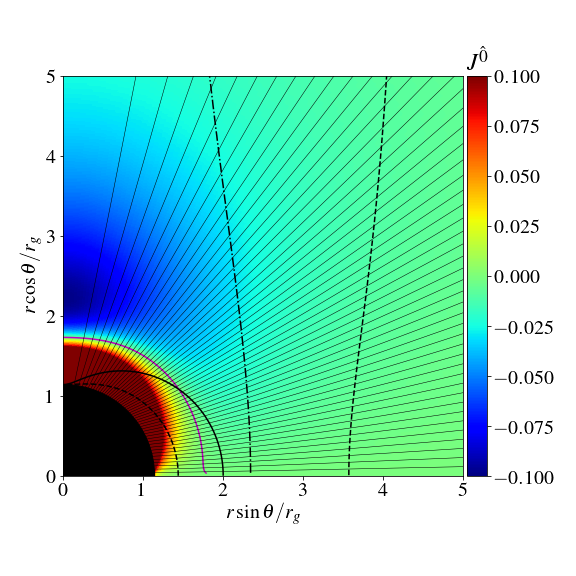}
\caption{Force-free split monopole solution around a Kerr black hole with spin $a=0.99$, in Boyer-Lindquist coordinates. The color map shows the charge density $J^{\hat{0}}$ as measured by ZAMO. Thin black lines---contours of the flux function $\Psi$; black solid line---ergosphere; black dashed lines---light surfaces; black dash-dotted line---stagnation surface; magenta line---$J^{\hat{0}}=0$. }\label{fig:FFsolution}
\end{figure}

Consider a nearly force-free, steady monopolar magnetosphere around a Kerr black
hole (Figure \ref{fig:FFsolution})\footnote{Most features we discuss are generic
  and apply in various magnetospheric configurations.}. Particles moving in the
strong magnetic field with Larmor radius $r_L\ll r_g\equiv GM/c^2$ slide along
the field line like beads on a wire. Because of the existence of two light
surfaces, particles are flung outward to infinity through the outer light
surface and flung inward towards the event horizon through the inner light
surface. In between there is a stagnation surface, located at the maximum of an
effective potential. The nature of the particle motion indicates that even if
the magnetosphere is initially filled with plasma, particles will inevitably
leak out from the two light surfaces. When the plasma density becomes too low to
conduct the current required by a force-free magnetosphere, the rotation induced
electric field will have a parallel component that cannot be screened, forming a
gap and accelerating particles to high enough energies, which then produce high
energy photons through inverse Compton (IC) or synchrotron/curvature processes
and initiate a pair cascade that replenishes the plasma, restoring the
magnetosphere to near force-free.

The charge and current densities required to maintain a force-free magnetosphere
have a few important properties for the monopole solution: (1) the poloidal
current is constant along the flux tube; (2) the 4-current is spacelike
everywhere; (3) there exists a null surface where the zero angular momentum
observer(ZAMO) measured charge density $J^{\hat{0}} = 0$ (Figure
\ref{fig:FFsolution}). The null surface has been regarded as a point of
separation of the plasma since if $J^{\hat{k}}$ has the same sign as the slope
of $J^{\hat{0}}$, and the plasma is charge-separated, then the current is
conducted by opposite charges moving away from the null surface, opening a
vacuum region where parallel electric field can grow
\citep[e.g.][]{1986ApJ...300..500C,1992SvA....36..642B,1998ApJ...497..563H}.
Meanwhile, the stagnation surface has also been considered to be a place for the
gap to form \citep{2015ApJ...809...97B}. Whether there is a preferred gap
location has yet to be tested using kinetic simulations.

The magnetospheric gap has also been invoked to explain the fast $\gamma$-ray
variability observed from some AGN \citep{2011ApJ...730..123L,
  2014Sci...346.1080A, 2015ApJ...809...97B, 2017ApJ...841...61A,
  2018ApJ...852..112K}. However, so far in the literature the gap physics has
been treated based on over-simplified vacuum models. In this work, we will
focus on the low luminosity regime (such that MeV photons from the disk are not
enough to produce the necessary charges), and study the microphysics and
dynamics of the gap using radiative PIC simulations.

\section{Numerical setup}
\label{sec:setup}

We would like to model the physical system described in section \ref{sec:intro}
using the simplest physics possible while capturing all the salient features,
namely the existence of a null surface, a stagnation point, and two light
surfaces. Spacetime correction to the equations of motion for particles and
fields are responsible for these effects, while the magnitude of these
corrections are actually small compared with the electromagnetic forces.
Therefore we model the flux tube using a flat space model while trying to
capture these features using different means.

In the flat spacetime model, we simply have
1D Maxwell equations \citep[e.g.,][]{2013ApJ...762...76C,2013MNRAS.429...20T}:
\begin{equation}
  \label{eq:1d-maxwell}
  \frac{\partial E_r}{\partial t} = 4\pi(j_B - j_r), \qquad
  \frac{\partial E_r}{\partial r} = 4\pi(\rho - \rhogj).
\end{equation}
where we take the background $j_B$ to be constant and use a
spatial profile of $\rhogj$ similar to the GR background charge density. See the
third row of Figure \ref{fig:gap} for the background charge and current density
profiles.

We include the effect of the light surfaces in 1D using a model inspired by the
light cylinder effect of a rotating neutron star. Consider the classic Michel
monopole solution \citep{1973ApJ...180L.133M}:
\begin{equation}
  \label{eq:michel-monopole}
  \boldsymbol{B} =  \frac{\mu}{r^2} \,\hat{\boldsymbol{r}} +
  B_{\phi} \,\hat{\boldsymbol{\phi}}, \quad B_{\phi} = -\frac{\Omega r \sin\theta}{c} B_r = \beta_{\phi}B_r.
\end{equation}
The field lines form an Archimedean spiral and become mostly toroidal outside
the light cylinder where $\Omega r\sin\theta = c$, or $\beta_{\phi} = -1$. The
field line rotates at an angular velocity of $\Omega$, and for any particles
outside the light cylinder, the corotation velocity is superluminal. The
particle compensates by sliding along the field line outwards; the total
  velocity is always less than $c$.

We can easily derive the equation for the 1D constrained motion of a particle
along a monopolar field line:
\begin{align}
  \label{eq:eom}
  \frac{dp_r}{dt} &= mc\frac{\beta_{\phi}\Omega\sin\theta}{\gamma}\left( \frac{\gamma - p_r/mc}{1 + \beta_{\phi}^2} \right)^2 + qE_r, \\
  \frac{dr}{dt} &= v_r =c \frac{p_r/\gamma mc + \beta_{\phi}^2}{1 + \beta_{\phi}^2},
\end{align}
where $p_r$ is the canonical momentum and 
\begin{equation}
\gamma=\sqrt{(p_r/mc)^2+1+\beta_{\phi}^2}
\end{equation}
is the Lorentz factor of the particle.
One can immediately see that no matter the value of $p_r$, $|p_r/\gamma
mc| < 1$, so when $\beta_{\phi}^2 \geq 1$, $v_r > 0$ and the particle is only
allowed to move in one direction. In our numerical simulations we model the
light surfaces in exactly the same way. We assume a profile for $\beta_{\phi}$
that is linear and antisymmetric across the center of the simulation box,
reaching $\pm 1$ at the light surfaces which are located at 0.05 and 0.95 of the
simulation box. This way the inner light surface is simply a mirror of
the outer light surface; the mid point of the box will be our ``stagnation
surface''.

We use a simplified version of the code \emph{Aperture} developed by the author
Alex Chen as a part of his PhD thesis \citep{chen-thesis}. The code only evolves
the 1D equations listed above, but keeping the charge-conserving current deposit
scheme proposed by \citet{2001CoPhC.135..144E}. This ensures that Gauss's law is
satisfied at all times if it is satisfied initially, so we only need to evolve
the first of equation \eqref{eq:1d-maxwell}. Comparison between background and
numerical values of $\rho$ and $j$ in the third row of Figure \ref{fig:gap}
confirms excellent charge conservation over time.

\subsection{Choice of units and scales}
\label{sec:units}

In our flat spacetime model we take the background current density $j_B$ to be
constant, which naturally defines a plasma frequency and skin depth:
\begin{equation}
  \omega_p = \sqrt{\frac{4\pi e j_B}{mc}}, \qquad \lambda_p = \frac{c}{\omega_p}.
\end{equation}
Thus we measure time using $\omega_p^{-1}$ and length using $\lambda_p$. Also
choosing $j_B$ as the unit of current, the electric field will be measured by
\begin{equation}
  E_0=\frac{4\pi j_B}{\omega_p},\qquad \frac{eE_0\lambda_p}{mc^2} = 1. 
\end{equation}
which simply means that $\tilde{E}=1$ corresponds to a voltage drop of $mc^2$
over a single $\lambda_p$, where the tilde denotes a dimensionless quantity. In
this set of units, naturally we have the unit of energy being $mc^2$ and
momentum being $mc$. We define pair multiplicity as $\mathcal{M} =
  (n_++n_-)ec/j_B$.

The profile of $\rhogj$ adds another parameter since it varies on the length
scale of $r_g$. The computational domain needs to accommodate several $r_g$
since we would like to include both light surfaces. For the physical systems we
are interested in, e.g. M87, $\lambda_p/r_g$ is on the order of $10^{-8}$ (Table
\ref{tab:parameters}). It is extremely difficult to have such scale
separation in a PIC simulation. Therefore we rescale this ratio, keeping
$r_g \gg \lambda_p$, and develop a semi-analytical model to infer what would
happen at physical parameters.

\subsection{Mechanism for pair production}
\label{sec:pairs}

The dominant mechanism for pair production in the black hole magnetosphere is
the collision of high energy photons with the low-energy photons from the disk.
The high energy photons come mostly from IC scattering of the background soft
photons by energetic leptons. We carry out the full radiative transfer including
IC scattering and the subsequent photon-photon collision assuming both happen on
the same background photon field. We assume a soft photon spectral distribution
of $I(\epsilon) = \epsilon dN/d\epsilon = I_0(\epsilon/\emin)^{-\alpha}$ which
cuts off at $\emin$ and extends up to $\sim0.1$ MeV. We use the Monte Carlo
method to sample the photon energy from a single IC event, then compute its free
path by drawing from an exponential distribution with a mean $\ellgg$. We track
this photon until it is converted to an $e^{\pm}$ pair at the end of its free
path. When the photon is not energetic enough to convert within the box, we do
not track it, but still cool the particle as if it emitted the photon.

The mean free path $\ellgg$ is energy dependent. The smallest mean free path
occurs when $\tilde{E}_\mathrm{ph}\temin \sim 2$ where $\ellgg \approx 5\ellic$,
and $\ellic$ is the characteristic IC mean free path in Thomson regime, $\ellic
= 1/n_s\sigma_T = \alpha/I_0\sigma_T$. For lower photon energy, $\ellgg$
increases:
\begin{equation}
  \ellgg(E_\mathrm{ph}) \approx 5\ellic\left( \frac{\tilde{E}_\mathrm{ph}\temin}{2} \right)^{-\alpha}
\end{equation}

The modeling of the IC process introduces several new numerical parameters: the
spectral index $\alpha$, the peak soft photon energy $\emin$, and a
characteristic free path for IC scattering $\ellic$. $\emin$ sets the energy
scale of the discharge, while $\ellic$ puts a new length scale into the problem.
The inferred characteristic values of these parameters are listed in Table
\ref{tab:parameters}. In our rescaling of the problem we focus on the
  optically thick regime, and ensure parameter ordering $\lambda_p\ll \ellic
\ll r_g$.

\begin{figure*}[t]
  \centering
  \hspace*{-0.7in}
  \includegraphics[width=1.15\textwidth]{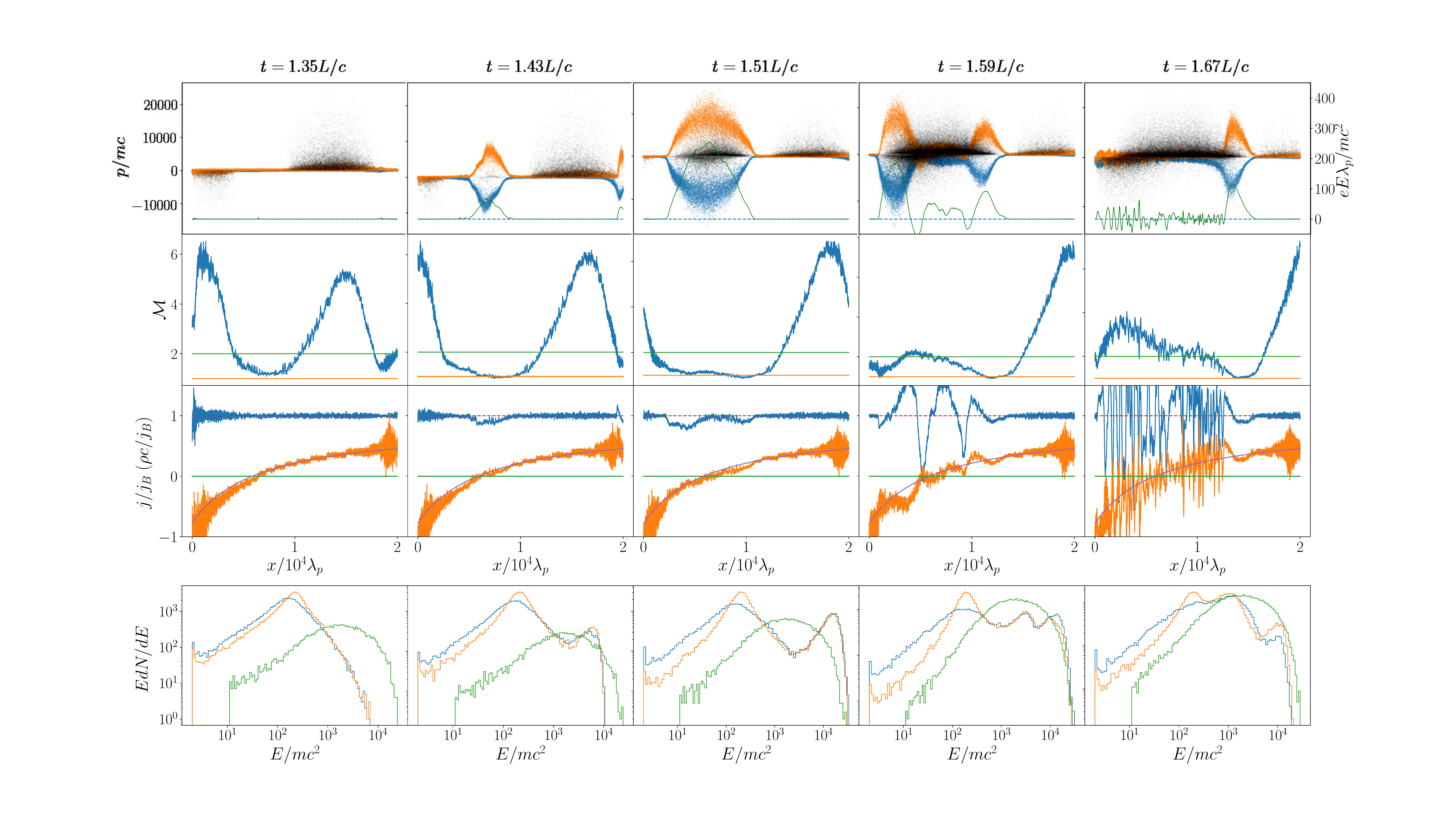}
  \vspace*{-0.5in}
  \caption{Time evolution of the gap. From left to right are snapshots at
    labeled times, where $L$ is the size of the box. The 4 panels from top to
    bottom are: 1) Phase space plots for electrons (blue), positrons (orange),
    and photons (black). The green line is electric field and its scale is on
    the right. 2) Pair multiplicity $\mathcal{M}$. Orange and green lines mark
    $\mathcal{M} = 1$ and $2$ respectively. 3) Current $j$ (blue) and charge
    density $\rho$ (orange) and their background values. 4) Spectrum of
    electrons (blue), positrons (orange), and photons (green).}
  \label{fig:gap}
\end{figure*}

\section{Time-dependent gap in 1D}

\subsection{Simulation results}
\label{sec:simulation}

We start from a plasma-filled initial condition where $E=0$ and $\rho = \rhogj$.
The initial pair multiplicity is 2, and all particles start at rest. Initially
small plasma-scale electric field develops to help the current to flow, but
since the box is leaky on both ends, plasma multiplicity drops over time. This
happens fastest where $d\rhogj/dx$ is largest. When $\mathcal{M} \lesssim 1$, an
electric gap opens locally to accelerate particles to high Lorentz factors, and
subsequently initiate pair creation, screening this gap, and launching
macroscopic bunches of pair plasma to both directions. Screening of the electric
field creates oscillations similar to those described by
\citet{2005ApJ...631..456L}. Eventually when the pairs are advected out of the
light surfaces and multiplicity drops again, the same cycle is initiated. In the
full length of one simulation, we are able to see several cycles of the gap
formation. We also tried starting with a vacuum electric field, but obtained the
same solution.

  Figure \ref{fig:gap} shows one such gap cycle, where we used $\alpha=1.2$,
  $\ellic=10\lambda_p$, $r_g/\lambda_p\sim 10^4$, and $\temin=10^{-6}$. It is
  the third time the gap develops in the simulation. As multiplicity drops from
  the previous cycle of pair creation, the system tries to maintain a
  macroscopic region as large as possible with $\mathcal{M}\gtrsim 1$ by drawing
  plasma from the side. When plasma flow can not sustain this state, a gap opens
  quickly over the whole region where $\mathcal{M}\sim 1$. As a result, in all
  our simulations the gap size $h \sim r_g$, and depends weakly on all
  parameters. The gap shown develops around the null surface, but it is not
    a guaranteed feature. It tends to develop wherever local multiplicity drops
    below unity, which can be anywhere due to plasma flow and delayed
    conversion of photons.

The photon spectrum shown in Figure \ref{fig:gap} is not to be confused with the
observable one. Due to limits of computational power, we only track photons that
convert to pairs within the box, so the shown spectrum should be interpreted
more as an absorption spectrum. In fact, most of the dissipated power in the gap
goes into radiation that leaves the box, only a small fraction of it converting
into $e^{\pm}$ pairs. The peak multiplicity from the gap is usually
$\mathcal{M}\sim 10$.

There are two well-defined spectral peaks for the particle energy shown in
Figure \ref{fig:gap}. When the gap is screened, the low energy peak is a
spectral break where the IC cooling becomes ineffective, $\ell_\mathrm{cool, IC}
\sim L$.
When the gap
is open, another spectral peak arises at higher energy. These are the primary
particles accelerated in the gap, and the peak energy $\gamma_p$ is controlled
by the gap electric field and IC cooling.

\subsection{Physics of the gap}
\label{sec:gap}

Consider a region in the magnetosphere
where plasma multiplicity $\mathcal{M}=1$ and $j=j_B$ at $t=0$. Electrons and positrons have to be counter streaming at speed of light to provide the current, so the number density of each species evolves as
\begin{equation}
  n_{\pm}(x, t) = \frac{j_B \pm \rhogj(x\mp ct)c}{2ec},
\end{equation}
Assuming $\rhogj$ varies linearly across this region with a slope $k =
d\rhogj/dx \sim j_B/r_gc$, we see that the current decreases over time if $k j_B>0$:
\begin{equation}
  j(t) = en_+(x, t)c + en_{-}(x, t)c \approx j_B-kc^2t.
\end{equation}
In particular, the time scale for the decrease of current depends on the
spatial scale over which $\rhogj$ varies. As a result, the electric field at the center of the gap increases as $t^2$
\begin{equation}\label{eq:E growth}
  E_{\parallel} =  \frac{1}{2}k c^2 t^2.
\end{equation}
The gap starts to be screened when enough photons emitted by the primary particles convert to pairs within the gap. 
During the characteristic time $t$, a primary particle goes through a number of 
\begin{equation}
N(\epsilon)=\frac{ct}{\ellic}\left(\frac{\epsilon}{\emin}\right)^{-\alpha}
\end{equation}
scatterings with target photons of energy $\epsilon$ ($\ge\emin$), generating $\gamma$-rays at energy $E_{\rm ph}=2\gamma_p^2\epsilon$ (applicable when $\gamma_p\tilde{\epsilon}\lesssim0.1$). 
Among these $\gamma$-rays, a number of $\kappa$ convert to pairs within time $t$:
\begin{equation}\label{eq:No. of pairs}
\kappa=N(\epsilon)\frac{ct}{\ellgg(E_{\rm ph})}=\frac{c^2t^2}{5\ellic^2}\left(\gamma_p^2\temin^2\right)^{\alpha},
\end{equation}
which turns out to be independent of $\epsilon$. For the gap to be screened,
$\kappa$ needs to be on the order of $1-10$ (we find that $\kappa\sim10$
reproduces well the results of our production runs). For most parameter regimes
of interest, when the screening happens the primary particles have short enough
cooling lengths such that the electrostatic acceleration
is balanced by the IC loss, so $\gamma_p$ is determined by $E_{\parallel}$
through
\begin{equation}\label{eq:balance}
e E_{\parallel}=\frac{4}{3}\gamma_p^2\frac{\alpha}{\alpha-1}\frac{\emin}{\ellic}.
\end{equation}
Using Equations (\ref{eq:E growth})(\ref{eq:No. of pairs})(\ref{eq:balance}), we can then obtain the peak electric field
\begin{equation}
  \label{eq:e-gap}
\frac{E_{\parallel}}{E_0}=\left(\frac{5 \kappa \ellic^2}{8 \pi \lambda _p r_g} \left(\frac{3 (\alpha -1)}{4 \alpha    }\frac{\ellic}{\lambda _p}\temin\right)^{-\alpha }\right)^{\frac{1}{\alpha +1}},
\end{equation}
and $\gamma_p$ can be calculated from Equation (\ref{eq:balance}). From
  $E_{\parallel}$, and gap size $h\sim r_g$, we can estimate the maximum gap
  power
  \begin{equation}
    \label{eq:gap-power}
    L_{\rm gap}\sim E_{\parallel}j_Br_g^3 \sim \frac{E_{\parallel}}{B}L_{\rm jet}.
  \end{equation}
We expect most of this power to be radiated away in gamma-rays.

\begin{figure}[h]
  \centering
  \includegraphics[width=\columnwidth]{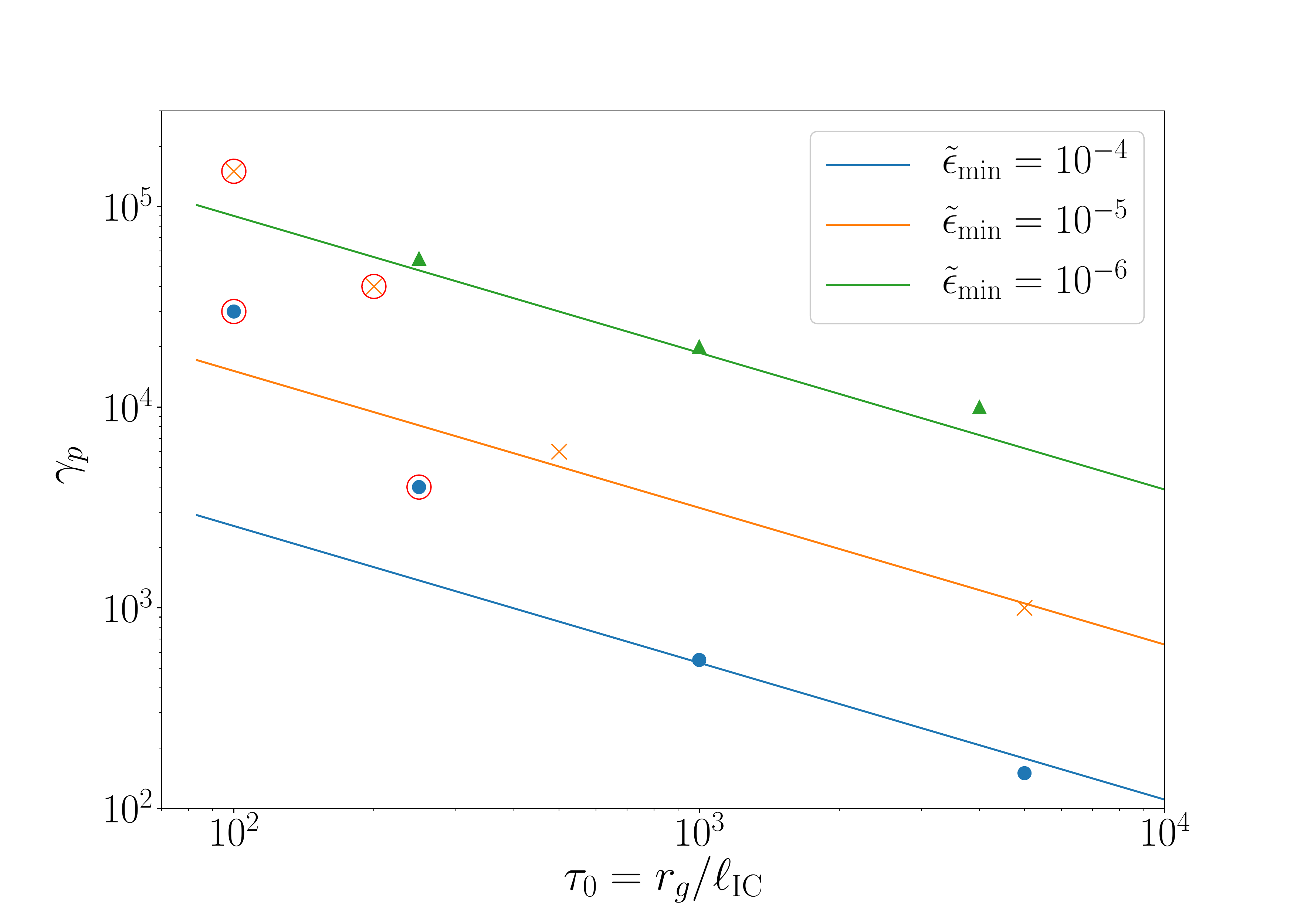}
  \caption{Scaling of primary particle Lorentz factor $\gamma_p$ with
      optical depth and soft photon peak $\emin$. Lines are our analytical
      predictions (equations \eqref{eq:balance} and \eqref{eq:e-gap}).
    Triangles, crosses, and dots are runs with $\temin=10^{-6}$, $10^{-5}$,
    $10^{-4}$ respectively, and should be compared with corresponding colored
    lines. All simulations are run with $\alpha=1.2$ and $r_g/\lambda_p=10^4$.
    The red-circled points are above $0.1/\temin$, close to Klein-Nishina
    regime, and our analytical model is no longer applicable.}
  \label{fig:scaling}
\end{figure}

Figure \ref{fig:scaling} shows that for all runs below the Klein-Nishina regime,
there is good agreement between the analytical model and the measured scaling
from the simulations. However, the above calculation no longer holds well if
primary particle energy approaches the Klein-Nishina regime:
$\gamma_p\gtrsim0.1/\temin$, which happens at relatively small optical depth
$\tau_0=r_g/\ellic\lesssim$ a few hundred. In that case, IC cooling becomes less
efficient and our argument for radiation balanced acceleration breaks down. We
expect $\gamma_p$ and the gap power to be much higher than our model would
predict, which is what we see in the simulations. The gaps in this regime tend
to be larger, but are still screened quasi-periodically as long as $\tau_0>$ a
few.

In the limit where $\ellic\gtrsim r_g$ or $\tau_0\lesssim1$, we found that it is increasingly
difficult to screen the gap, which develops to encompass the whole domain.
Particles are accelerated into deep Klein-Nishina regime where $\gamma_p\gg
1/\temin$, and $E_{\parallel}\gtrsim B$. In this limit we expect significant
changes of the magnetospheric structure due to the gap, possibly killing the
  jet structure, and 1D approximation we employed in this paper is no longer
appropriate.

\subsection{Scaling to real systems}
\label{sec:application}

\begin{deluxetable}{c|cc}
\tablecaption{Parameters for M87 and Sgr A* \label{tab:parameters}}
\tablehead{
 &
\colhead{M87} &
\colhead{Sgr A*}
}
\startdata
$M\; (M_{\odot})$ & $6.6\times10^9$ & $(4.3\pm0.4)\times10^6$\\
$r_g\; ({\rm cm})$ & $10^{15}$ & $6.4\times10^{11}$\\
$L_s/L_{\rm Edd}$\tablenotemark{a} & $10^{-6}$ & $10^{-9}$\\
$B$ (G)\tablenotemark{b}  & 200 & 30 \\
$n_{\rm GJ}\; ({\rm cm}^{-3})$ & $5\times10^{-4}$ & $0.1$\\
$\lambda_p/r_g$ & $2\times 10^{-8}$ & $2.6\times10^{-6}$\\
$u_s\; ({\rm erg\, cm^{-3}})$\tablenotemark{a}& 0.1 & 0.15\\
$n_s\; ({\rm cm^{-3}})$\tablenotemark{a} & $10^{13}$ & $2\times10^{13}$\\
$\emin$ (meV) & 1.2 & 1.2\\
$\emin/m_ec^2$ & $2.3\times10^{-9}$ & $2.3\times10^{-9}$\\
$\alpha$ & 1.2 & 1.25\\
$\ellic/r_g$ & $1.5\times10^{-4}$ & $0.12$\\
\hline
$E_{\parallel}/E_0$ & $1.8\times10^3$ & $2.9\times10^4$ \\
$E_{\parallel}/B$ & $3.6\times10^{-5}$ & 0.075\\
$\gamma_p$ & $2.7\times10^7$ & $3\times10^8$\\
$L_{\rm gap}\; ({\rm erg\,s^{-1}})$ & $3.6\times10^{39}$ & $4\times10^{34}$\\
\enddata
\tablenotetext{\textrm{a}}{$L_s$, $u_s$, and $n_s$ are soft photon luminosity, energy
  density, and number density, at a few ($\sim 3$) $r_g$ from the black hole: $u_s=L_s/(4\pi r_g^2c10)$.}
\tablenotetext{\textrm{b}}{The poloidal magnetic field near the event horizon, estimated based on the jet power $L_{\rm jet}\sim a^2 c B^2r_g^2/4\pi$. For M87, $L_{\rm jet}\sim 10^{44}\, {\rm erg\, s^{-1}}$; for Sgr A*, we assume $L_{\rm jet}\sim L_s$.}

\end{deluxetable}

The most relevant systems where the spark gap might exist are low luminosity
AGN like M87 and Sgr A*. We list the physical parameters inferred from
observation in Table \ref{tab:parameters}, as well as predictions from our
physical model. The observational parameters are based on
\citet{2015ApJ...809...97B}. For M87 we expect it to be well described by our
model, and indeed the predicted $\gamma_p \sim 3\times 10^7$ is well below the
KN regime. The typical gamma-ray photons that are produced by these primary
particles will be in the range of $0.1$ to a few TeV; most of them will escape
the outer light surface. This coincides with the observed energy range of the TeV
flares from M87 \citep{2012ApJ...746..151A}. Our time-dependent gap model also
predicts time variability of several $r_g/c$, which for M87 would be about
$\sim$day, again coinciding with the observed time scale of the flares. However,
the total gap power predicted by our model is at best only consistent with the
quiescent state, and too low for the flares. Whether this mechanism can explain
the origin of M87 flares will be investigated in a future paper.

For Sgr A* however, $\gamma_p>0.1/\temin$, and we are in the Klein-Nishina
regime. In this case we expect the actual primary particle energy to be higher,
and $L_\mathrm{gap}$ might become comparable to $L_\mathrm{jet}$. As a result,
the simplistic 1D approximation we adopted in this paper is no longer
applicable, as this gap should be able to significantly affect the global
magnetosphere structure. This potentially can explain the lack of an
  apparent jet structure from the center of our galaxy. To properly treat this
regime a global magnetospheric simulation will be needed.

\section{Discussion}

We have presented self-consistent 1D simulations of pair cascade in a magnetized
plasma within the black hole magnetosphere. Informed by the numerical results we
developed a semi-analytical model for the electric gap, providing an estimate for
gap power in systems that are optically thick to inverse Compton scattering.

Traditionally the study of the discharge problem in the BH magnetosphere were
often based on a vacuum gap model around the null surface, drawing analogy to
the outer gap model in pulsar magnetospheres \citep[e.g.][]{2016A&A...593A...8P,
  2017arXiv170600542F}. We have shown through numerical simulations that the
physical conditions for such models are never realized: the domain never
tolerates a local vacuum region, nor a static gap. Electric field develops
to accelerate leptons as soon as the local multiplicity drops below unity,
initiating the process of pair discharge. Moreover, the gap can develop anywhere
depending on plasma flow, not necessarily at the null surface.

We did not include the GR correction to the particle equations of motion.
Instead, the GR effect is entirely captured by the varying background charge
density $\rhogj$, and the presence of an inner light surface. Without GR both
features will be absent. We think this is an appropriate simplification that
allows us to focus on the electrodynamics and microphysics. A proper general
relativistic set of equations could in principle be implemented, as was recently
done by \citet{2018arXiv180304427L}. However, they report an overall
quasi-steady state which is different from what we observe. We believe the main
differences are the treatment of light surfaces, and they focus on a low optical
depth regime $r_g/\ellic \lesssim 10$. The logical extension of the results in
this paper is to look at how the global structure of the magnetosphere will
interact with the gap, especially when the gap power becomes comparable to the
jet power.

\acknowledgments We thank Matthew Kunz, and Alexander Tchekhovskoy for very
helpful discussions, and thank Andrei Beloborodov, Anatoly Spitkovsky for their
constructive comments on the manuscript. YY acknowledges support from the Lyman
Spitzer, Jr. Postdoctoral Fellowship awarded by the Department of Astrophysical
Sciences at Princeton University. AC acknowledges the support of NASA grant
NNX15AM30G. YH is supported by NSERC and in part by the Perimeter Institute for
Theoretical Physics. Research at Perimeter Institute is supported by the
Government of Canada through the Department of Innovation, Science and Economic
Development Canada and by the Province of Ontario through the Ministry of
Research, Innovation and Science. The code used in this paper is available at
\url{https://github.com/fizban007/1Dpic} under GPLv2.0.

% \bibliographystyle{yahapj}
% \bibliography{ref} 

\end{document}